\documentclass[runningheads]{llncs}
\usepackage[T1]{fontenc}
\usepackage{graphicx}
\usepackage{multirow}
\usepackage{tabularx}
\usepackage{makecell}
\usepackage{listings}
\usepackage{siunitx}
\usepackage[hidelinks]{hyperref}
\usepackage{xcolor}
\colorlet{punct}{red!60!black}
\definecolor{background}{HTML}{EEEEEE}
\definecolor{delim}{RGB}{20,105,176}
\colorlet{numb}{magenta!60!black}
\lstdefinelanguage{json}{
    basicstyle=\footnotesize\ttfamily,
    numbers=left,
    numberstyle=\tiny,
    stepnumber=1,
    numbersep=8pt,
    showstringspaces=false,
    breaklines=true,
    frame=lines,
    backgroundcolor=\color{background},
    literate=
     *{0}{{{\color{numb}0}}}{1}
      {1}{{{\color{numb}1}}}{1}
      {2}{{{\color{numb}2}}}{1}
      {3}{{{\color{numb}3}}}{1}
      {4}{{{\color{numb}4}}}{1}
      {5}{{{\color{numb}5}}}{1}
      {6}{{{\color{numb}6}}}{1}
      {7}{{{\color{numb}7}}}{1}
      {8}{{{\color{numb}8}}}{1}
      {9}{{{\color{numb}9}}}{1}
      {:}{{{\color{punct}{:}}}}{1}
      {,}{{{\color{punct}{,}}}}{1}
      {\{}{{{\color{delim}{\{}}}}{1}
      {\}}{{{\color{delim}{\}}}}}{1}
      {[}{{{\color{delim}{[}}}}{1}
      {]}{{{\color{delim}{]}}}}{1},
      morestring=[b]"
}
\usepackage[inline]{enumitem}
\begin{document}
\title{A modular and scalable web platform for computational phylogenetics}
%
%
\author{Nyckollas Brandão\inst{1} \and
André Jesus \inst{1} \and
André Páscoa\inst{1} \and
Alexandre P. Francisco \inst{2,3} \and
Mário Ramirez \inst{4} 
Cátia Vaz \inst{1,3}
}
\authorrunning{N. Brandão et al.}
%
\institute{Instituto Superior de Engenharia de Lisboa, Instituto Politécnico de Lisboa \and
Instituto Superior Técnico, Universidade de Lisboa \and
INESC-ID Lisboa
   \and
Instituto de Microbiologia, Instituto de Medicina Molecular, Faculdade de Medicina, Universidade de Lisboa
}
\maketitle              
\begin{abstract}
Phylogenetic analysis, which allow to understand the evolution of bacterial and viral epidemics, 
requires large quantities of data to be analysed and processed for knowledge extraction.
 One of the major challenges consists
on the integration of the results from typing and phylogenetic inference
methods with epidemiological data, namely in what concerns their integrated and simultaneous analysis and visualization. 
Numerous approaches to support phylogenetic analysis have been proposed, varying from standalone tools to integrative web applications that include tools and/or algorithms for executing the common analysis tasks for this kind of data. However, most of them lack the capacity to integrate epidemiological data. Others provide the ability for visualizing and analyzing such data, allowing the integration of epidemiological data but they do not scale for large data analysis and visualization. Namely, most of them run inference and/or visualization optimization tasks on the client side, which becomes often unfeasible for large amounts of data, usually implying transferring data from existing databases in order to be analysed. Moreover, the results and optimizations are not stored for reuse.
We propose the PHYLOViZ Web Platform, a cloud based tool for phylogenetic analysis, that not only unifies the features of both existing versions of PHYLOViZ, but also supports structured and customized workflows for executing data processing and analyses tasks, and promotes the reproducibility of previous phylogenetic analyses. This platform supports large scale analyses by relying on a workflow system that enables the distribution of parallel computations on cloud and HPC environments. Moreover, it has a
modular architecture,  allowing easy integration of new methods and tools, as well as customized workflows, making it flexible and extensible.

\end{abstract}
\section{Introduction}\label{cap:problem-description}
Phylogenetic analysis raises several computational challenges.
Depending on the tools, type and amount of data, some steps can take a long time to be executed.
For instance, the alignment of sequences or the application of inference algorithms over large datasets can be time-consuming and require non-trivial computational resources.
Another problem is the efficiency of the visualization process, since it involves processing and rendering large amounts of data, which can be unfeasible to be executed in its entirety on the client side.
Finally, the management of multiple datasets and associated results becomes also challenging in this context, namely when reproducibility is of utmost importance.

The phylogenetic analysis computational methods and tools are often subject to high levels of
customization, including:
\begin{enumerate*}[label=(\roman*)]
    \item algorithms and tools selection as well as their parameterization,
    \item data and tree visualization layouts choice, and
    \item ancillary data selection and integration, namely within the visualization.
\end{enumerate*}
On one hand this choices must be tracked and stored together with the results.
On the other hand modularity is fundamental in order to sort different and new methods and algorithms.
Modularity is an important feature for a software system, since it allows for the easy integration of new modules and the improvement of existing ones, making it flexible and extensible~\cite{richards2020fundamentals}.

With this in mind, a platform should address the above mentioned challenges by attending to the following objectives:
\begin{enumerate*}[label=(\roman*)]
    \item have a user friendly platform that allows users to access, manage and perform phylogenetic analyses and visualizations easily and intuitively, preserving data and results for future use;
    \item provide a variety of customization options, including the choice of what algorithms and tools, as well as their parameterizations are used for the analyses;
    \item rely on state of the art workflow management systems for the integration of multiple and complex phylogenetic analysis workflows;
    \item support efficient and scalable visualizations of phylogenetic data and trees;
    \item store and manage phylogenetic data and analyses results efficiently;
    \item provide a modular architecture that allows for the easy integration of new analysis methods and tools.
\end{enumerate*}

In this paper we introduce the PHYLOViZ Web platform which aims to achieve these objectives. We start by providing some background on the phylogenetic context as well as on related alterantive tools and platforms. We present then the overall architecture, the underlying data model and platform use cases.
Implementation details are provided next. Finally we discuss its main characteristics and potential future work.

\section{Background}
Phylogenetic analysis \cite{huson2010phylogenetic} seeks to reveal the evolutionary relationships between various species, or even individuals or strains within the same species, in order to gain insights into their evolutionary history. The outcome of this analysis is a phylogeny, typically represented as a phylogenetic tree or network.

Generally, this analysis includes the following 4 steps. 
The first step is genetic sequence alignment, where genetic sequences from different isolates of a particular organism or pathogen are compared and arranged. Isolates are individual samples or strains obtained from various sources, such as patients or environmental samples.

The next step involves the application of a typing methodology~\cite{butler2011advanced}, which types or classifies sequences based on relevant genetic markers or characteristics. This process generates typing data, such as allelic profiles or nucleotide sequences, using methods like Multi Locus Sequence Typing (MLST)~\cite{artc:MLST,artc:MLST2} or Multi Locus Variable Number Tandem Repeat Analysis (MLVA)~\cite{artc:mlva}.

Following the typing process is the execution of a phylogenetic inference method~\cite{saitou2013introduction}, which generates a diagrammatic hypothesis of the evolutionary relationships within the group of organisms.
There are several types of phylogenetic inference methods, which may require the calculation of a distance matrix~\cite{vaz2021distance}. The resulting hypothesis is typically represented as a graph or a tree.

Finally, the analysis concludes with the visualization and exploration of inferred results. This step integrates epidemiological and ancillary data, such as demographic and clinical information, related to the isolates under study. This additional data provides further context and enhances the interpretation of the phylogenetic analysis results.

Several solutions for supporting phylogenetic analyses have been proposed, varying from standalone tools to integrative web applications that include tools and algorithms for common analysis tasks.

PHYLOViZ~\cite{francisco2012phyloviz,nascimento2017phyloviz} is a Java desktop application. It supports multiple customization options, including the choice of different inference algorithms, layouts and transformations on tree visualizations, as well as data integration.
It is also plugin based, supporting the addition of new algorithms and functionalities.
However, since it is a desktop application, all computations and data management occur on the client-side, and all data and results are stored only locally.
Furthermore, tree visualizations do not scale effectively for large datasets.

PHYLOViZ Online \cite{ribeiro2016phyloviz,phyloviz_online} was created as the web version of PHYLOViZ, so there is no need to download software.
It provides tree visualizations and data integration, although it only supports a single layout, namely the force-directed layout. The visualization is also not scalable for large datasets. Moreover, only a single inference algorithm is provided and, since it lacks modularity and it is not extensible, it is not easy to add new algorithms and functionalities.

The integration of results obtained from inference algorithms with epidemiological data (also called isolate or ancillary data) and simultaneous analysis is still limited by visualization and processing techniques. Although, besides PHYLOViZ and PHYLOViZ Online, there are other known tools for visualizing and analyzing such data, allowing the integration of epidemiological data.

SplitsTree~\cite{10.1093/molbev/msj030} is a widely used single user desktop application for computing unrooted phylogenetic networks from molecular sequence data.
Given an alignment of sequences, a distance matrix or a set of trees, the tool will compute a phylogenetic tree or network using methods such as split decomposition, neighbor-net, consensus network, super networks methods or methods for computing hybridization or simple recombination networks. The current release is SplitsTree4.

Phylogeny.fr \cite{Phylogeny.fr} is a free, simple to use web service dedicated to reconstructing and analysing phylogenetic relationships between molecular sequences.
Phylogeny.fr runs and connects various bioinformatics programs to reconstruct a phylogenetic tree from a set of sequences. There is also an upgraded version of Phylogeny.fr, called NGPhylogeny.fr, that supports execution of tasks in the Galaxy workflow system.
NGPhylogeny.fr~\cite{10.1093/nar/gkz303} is a webservice dedicated to phylogenetic analysis.
It provides a complete set of phylogenetic tools and workflows adapted to various contexts and various levels of user expertise.
It is built around the main steps of most phylogenetic analyses
However, both Phylogeny.fr and NGPhylogeny.fr have some limitations. Both of them do not allow the choice of the tree visualization layout, providing a single choice. Another limitation is that these services do not support multiple deployments and, even assuming that they are modular, their extensibility is limited to what the service maintainers decide to integrate.

GrapeTree~\cite{GrapeTree} supports the analyses of large datasets containing thousands of profiles.
GrapeTree is a stand-alone package for investigating phylogenetic trees plus associated metadata, being also integrated in EnteroBase~\cite{zhou2020enterobase} to facilitate cutting edge navigation of genomic relationships among bacterial pathogens.
The novelty is an inference algorithm that can deal with missing data.
It supports however a single inference algorithm and single visualization layout, and it is not easily extensible.
We note that the static visualization layout works for large datasets.

In what concerns the building blocks for this kind of platforms, and beyond tools implementing inference algorithms and providing visualizations, we highlight the need of workflow integration middlewares and suitable data models and storage engines.
In this work we will rely on the following components.
\begin{enumerate*}[label=(\roman*)]
\item FLOWViZ \cite{luis_2022_flowviz} is a middleware that allows to seamlessly integrate phylogenetic platforms with data-centric workflow scheduling and execution, namely through the Apache Airflow workflow system, although other workflow systems can be added.
\item PhyloDB~\cite{lourenco_2021_phylodb} is a web service framework designed for the efficient storage and management of large phylogenetic datasets; it is a database built on top of the graph database Neo4j, with a data model specific for phylogenetic data, which allows for fast querying. 
\item PhyloLib~\cite{silva_2020_library} is a library of efficient algorithms for phylogenetic analysis that can be easily containerized and integrated in workflows.
\end{enumerate*}

\section{PHYLOViZ Web Platform} \label{sec:solution}
The proposed solution is a modular and extensible platform easily deployable in cloud environments and integrable with Apache Airflow, a data-centric workflow system. It supports multiple users that can manage multiple datasets and related results, as well as phylogenetic trees visualizations and ancillary data integration. Phylogenetic analyses are abstracted and implemented as self-contained workflows and containerized tools.
The PHYLOViZ Web Platform is available at \url{https://github.com/phyloviz/phyloviz-web-platform}, including detailed documentation for users and developers.

\begin{figure}[t]
    \centering
    \includegraphics[width=\textwidth]{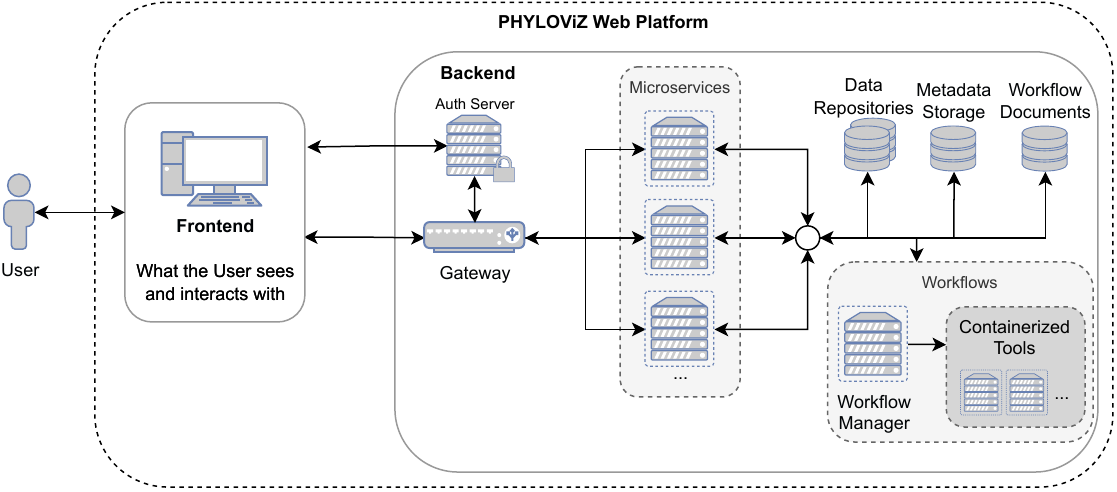}
    \caption{PHYLOViZ Web Platform architecture.} \label{fig:architecture}
\end{figure}

The architecture of the platform is depicted in Figure~\ref{fig:architecture}.
It consists of a backend application (server-side) and a frontend application (client-side).
The backend comprises a collection of microservices~\cite{li2021understanding}, responsible for data processing and storage. It relies on workflows and containerized tools to execute required analyses.
The frontend is composed of a web application, that facilitates data visualization and user interaction. Communication between the frontend and the backend is achieved through a REST API available through a gateway.

\begin{figure}[t]
    \centering
    \includegraphics[width=\textwidth]{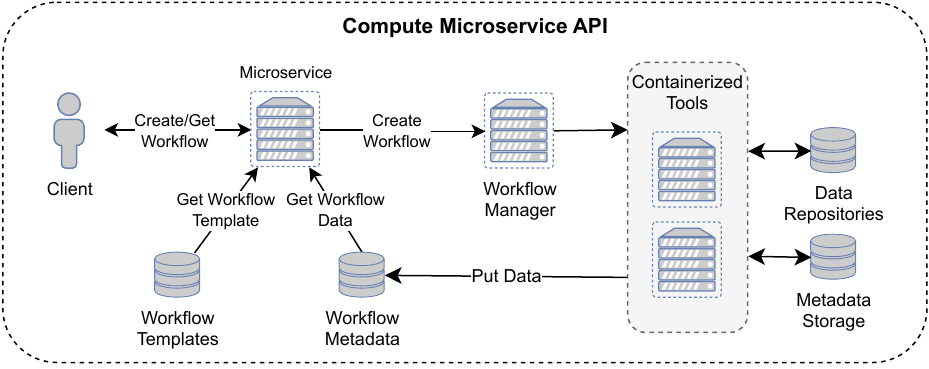}
    \caption{Compute microservice architecture.} \label{fig:compute-microservice-architecture}
\end{figure}

The microservices communicate with the data stores and repositories, and underlying services.
We have the following microservices:
\begin{enumerate*}[label=(\roman*)]
    \item {\em administration}, responsible for projects and datasets management;
    \item {\em file transfer}, responsible for the upload and download of project files;
    \item {\em compute}, responsible for workflows execution and management;
    \item {\em visualization}, responsible for the visualization of results and underlying data integration.
\end{enumerate*}
The {\em compute} microservice is depicted in Figure~\ref{fig:compute-microservice-architecture}
for illustrating the microservice architecture adopted in the proposed solution.

The underlying data model splits data into two categories: metadata and phylogenetic data, as depicted already in Figure~\ref{fig:architecture}. In addition to these categories, it also stores documents related to workflows, including templates, configurations and instances ready to be executed.
Metadata describes the application resources, namely users, projects, datasets, results, and visualizations, including their properties, relationships and dependencies.
Additionally, metadata serves as a reference to the data representations of these resources stored within the data repositories, which contain the actual content associated with each resource.
For instance, inferred phylogenetic trees are stored exclusively as a newick formatted file in the designated S3 bucket, the data repository. But associated metadata is also stored including information about the related project and dataset, the repository where it is located, as well as which workflow was used to generate it.
Metadata is stored in MongoDB.
Phyogenetic data is also indexed in a storage engine, PhyloDB, for fast data querying and processing.

It is important to note that, in this context, a dataset refers to a phylogenetic study and as such contains the data relevant to that study, as well as analyses results, including multiple inferred distance matrices, trees and visualizations.
Each dataset within a project is separate from each other, and operates on its own set of resources and data, even though the typing data and isolate data files of each dataset can be shared among all the datasets in the project. This allows for flexibility and customization of each dataset without duplicating data.
Analyses results and visualizations are not shared between datasets.

\begin{figure}[t]
    \centering
    \includegraphics[scale=0.54]{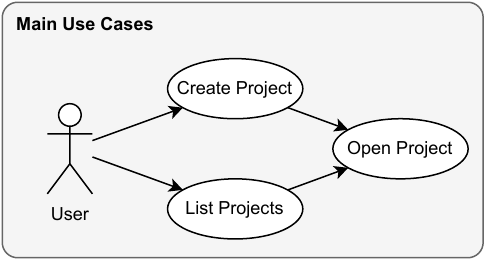}
    \includegraphics[scale=0.54]{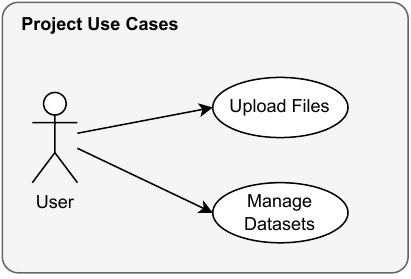}
    \includegraphics[scale=0.54]{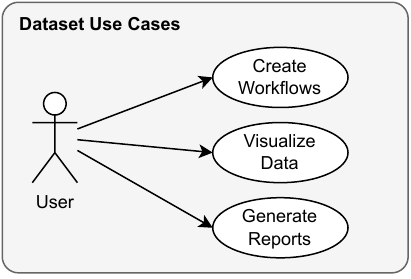}
    \caption{PHYLOViZ Web Plataform use cases.}
    \label{fig:main-use-cases}
\end{figure}

The platform use cases are divided into three categories: main use cases, project use cases and dataset use cases (see Figure~\ref{fig:main-use-cases}).
The main use cases are the ones that are available when the user accesses the platform. 
The project use cases are the ones that are available when a project is open. The user has access to the project information, including files and datasets. 
A new dataset can be created or an existing dataset can be selected. Opening a dataset provides access to the previously saved state and associated information, with the user being able to seamlessly continue analyses from where they were left off.
The user can also upload files to the project, that can be used to create datasets.
The dataset use cases are the ones that are available when a dataset is selected. The user can perform phylogenetic analysis tasks through the execution of workflows, which will be executed in the background and the user notified upon its termination.
The user can also visualize the results of these tasks and generate reports.

The extensibility of the PHYLOViZ Web Platform is a fundamental aspect of its design, allowing administrators and developers to customize and expand the platform according to their specific requirements. Let us detail two key elements of extensibility: creating and customizing workflows, and establishing data repositories. 
By offering the flexibility to define workflows and effectively manage data, the platform becomes a versatile tool that can adapt to various research needs. Through these extensibility features, researchers can seamlessly integrate their analysis pipelines, incorporate new analysis methods, and organize and access their phylogenetic data. 

Customizing workflows, whether it involves adding, removing, or editing elements, requires configuring:
\begin{enumerate*}[label=(\roman*)]
    \item {\em workflow templates} that define the tasks executed, the corresponding commands to be run, and the arguments received by each task;
    \item {\em tool templates} that specify the Docker image associated with each tool and other tool-related configurations and parameterizations; and
    \item auxiliary {\em tools} that may be required to seamlessly integrate workflows into the application, e.g., to retrieve inputs and to deliver results.
\end{enumerate*}

\section{Implementation}\label{cap:implementation}

\begin{figure}[t]
    \centering
    \includegraphics[width=1.0\textwidth,keepaspectratio]{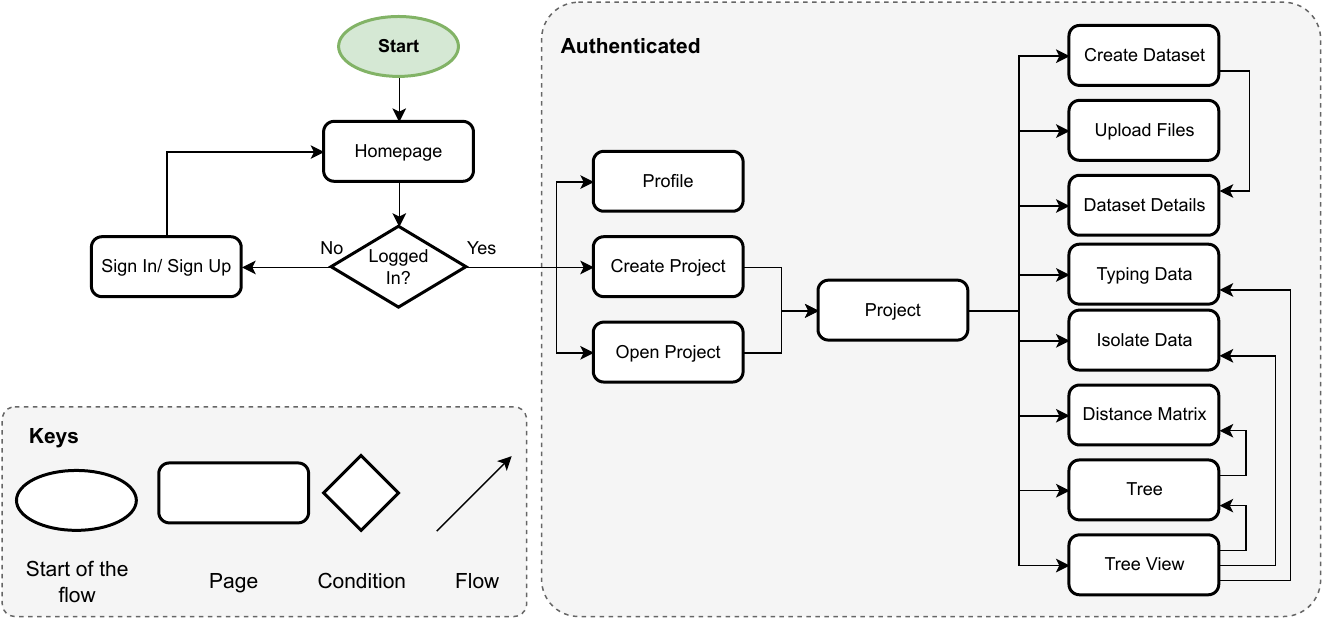}
    \caption{UI flow diagram.} \label{fig:ui-flow-diagram}
\end{figure}

Let us describe the main implementation choices.
The frontend application is a web application through which the user interacts with the platform.
This application provides a simple and intuitive interface allowing the user to manage projects and datasets, upload files, execute computation tasks, and visualize the results, among other functionalities.
The user interface (UI) flow diagram is depicted in Figure~\ref{fig:ui-flow-diagram}.
This web application is implemented using React with TypeScript. 

The services in the frontend application are responsible for communicating with the backend application. Each backend microservice has a corresponding service in the frontend application, which acts as a namespace containing functions responsible for handling each request to the backend. These functions are asynchronous and return a promise that resolves to the response of the request.
To facilitate communication with the backend, the Fetch API~\cite{fetch} is used, which is a modern replacement for the older \verb|XMLHttpRequest|. The Fetch API provides client functionality for transferring data between a client and a server, and is known for its simplicity and cleanliness, using promises instead of callbacks.

\begin{figure}[t]
    \centering
    \includegraphics[width=1.0\textwidth,keepaspectratio]{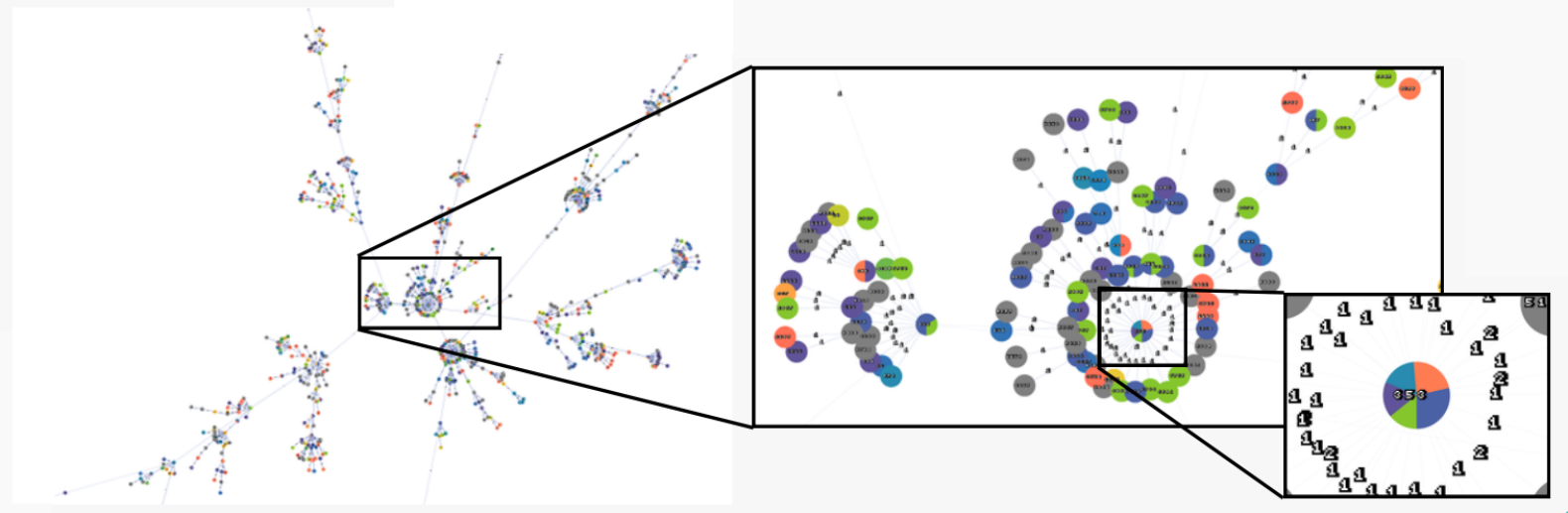}
    \caption{Tree visualization details.} \label{fig:ui-force-direct-layout}
\end{figure}

The component for visualize trees is one of the most complex components in the application. It is responsible for the tree visualization, including rendering and application of transformations to the visualization. It relies in many visual components and libraries to achieve its functionality. One notable library is \verb|cosmos|, used for the force-directed layout. It enables the manipulation and transformation of the tree visualization, providing features such as zooming, panning, and applying various graph algorithms for layout and organization.
In addition to the \verb|cosmos| library, this component also leverages the functionality provided by the \verb|chart.js| library, which provides charting capabilities for rendering the tree visualization. 
Several enhancements were implemented in the \verb|cosmos| library. These improvements include the implementation of node and link labels, node dragging, and the integration of pie charts on the nodes.
One of the main challenges faced was the need for efficient rendering, especially when visualizing thousands of nodes. To address this, a batching technique was used to render all elements in a single draw call. This optimization allowed for smooth performance and improved the user experience when exploring large tree structures.
Figure~\ref{fig:ui-force-direct-layout} highlights the details in the tree visualization.

Let us briefly present the implementation of the backend microservices.
All microservices are implemented using Spring Boot, a framework that simplifies the development of stand-alone, production-grade Spring-based applications. Spring Boot offers a wide range of features and integrations that facilitate the creation of robust and scalable microservices.
The microservices in our platform share a common implementation pattern, ensuring consistency and ease of maintenance across the different components. This common implementation pattern is based on a layered architecture~\cite{richards2015software}, comprising the HTTP layer, Services layer, and Repository layer.
Components within this architecture pattern are organized into horizontal layers, each layer performing a specific role within the application (e.g., presentation logic or business logic). Although the layered architecture pattern does not specify the number and types of layers that must exist in the pattern, most layered architectures consist of four standard layers: presentation, business, persistence, and database.
By adhering to this layered architecture pattern, the microservices in our system exhibit modularity, maintainability, and scalability. The separation of concerns provided by this pattern enables easier testing, debugging, and extension of the system functionalities. It also allows for independent development and deployment of each microservice, fostering a more flexible and resilient architecture.
As mentioned earlier, platform microservices make their functionality available through a REST API. Table~\ref{tab:administration-endpoints-projects} provides a comprehensive overview of the available API endpoints and their corresponding operations.
\begin{table}[tp]
\centering
\caption{REST API endpoints provided by the platform microservices.}
\label{tab:administration-endpoints-projects}
\begin{tabularx}{\textwidth}{|p{20em}|p{46px}|X|}
    \hline
    \textbf{Endpoint} & \textbf{HTTP Method} & \textbf{Description} \\
    \hline
    /projects & GET & Retrieve a list of all projects \\
    \hline
    /projects & POST & Create a new project \\
    \hline
    /projects/\{projectId\} & GET & Retrieve details of a specific project \\
    \hline
    /projects/\{projectId\} & PATCH & Update a specific project \\
    \hline
    /projects/\{projectId\} & DELETE & Delete a project \\
    \hline
    /projects/\{projectId\}/datasets & POST & Create a new dataset \\
    \hline
    /projects/\{projectId\}/datasets/\{datasetId\} & GET & Retrieve details of a specific dataset \\
    \hline
    /projects/\{projectId\}/datasets/\{datasetId\} & PATCH & Update a specific dataset \\
    \hline
    /projects/\{projectId\}/datasets/\{datasetId\} & DELETE & Delete a dataset \\
    \hline
    /projects/\{projectId\}/files/typing-data & POST & Upload typing data file \\
    \hline
    /projects/\{projectId\}/files/typing-data/ \{typingDataId\}/file & GET & Download typing data file \\
    \hline
    /projects/\{projectId\}/files/isolate-data & POST & Upload isolate data file \\
    \hline
    \makecell[l]{/projects/\{projectId\}/files/isolate-data/\\\{isolateDataId\}/file} & GET & Download isolate data \\
    \hline 
    /projects/\{projectId\}/workflows & GET & Retrieve workflows for a specific project \\
    \hline
    /projects/\{projectId\}/workflows & POST & Create a new workflow for a specific project \\
    \hline
    \makecell[l]{/projects/\{projectId\}/workflows/\\\{workflowId\}} & GET & Retrieve a specific workflow \\
    \hline
    \makecell[l]{/projects/\{projectId\}/workflows/\\\{workflowId\}/status} & GET & Retrieve the status of a specific workflow \\
    \hline
    \makecell[l]{/projects/\{projectId\}/datasets/\{datasetId\}/\\trees/\{treeId\}} & GET & Retrieve a tree \\
    \hline
    \makecell[l]{/projects/\{projectId\}/datasets/\{datasetId\}/\\tree-views/\{treeViewId\}} & GET & Retrieve a tree view \\
    \hline
    \makecell[l]{/projects/\{projectId\}/datasets/\{datasetId\}/\\distance-matrices/\{distanceMatrixId\}} & GET & Retrieve a distance matrix \\
    \hline
    \makecell[l]{/projects/\{projectId\}/files/isolate-data/\\\{isolateDataId\}/keys} & GET & Retrieve isolate data keys \\
    \hline
    \makecell[l]{/projects/\{projectId\}/files/isolate-data/\\\{isolateDataId\}/rows} & GET & Retrieve isolate data rows \\
    \hline
    \makecell[l]{/projects/\{projectId\}/files/typing-data/\\\{typingDataId\}/schema} & GET & Retrieve the typing data schema \\
    \hline
    \makecell[l]{/projects/\{projectId\}/files/typing-data/\\\{typingDataId\}/profiles} & GET & Retrieve the typing data profiles \\
    \hline 
    \end{tabularx}
\end{table}

Access management is implemented using Keycloak~\cite{thorgersen_2023_keycloak}, an open-source Identity and Access Management (IAM) solution that provides features such as authentication, authorization, and user management. It supports various protocols such as OAuth2, OpenID Connect, and SAML for authentication and authorization. In the context of PHYLOViZ Web Platform, Keycloak acts as an OpenID identity provider that authenticates users and issues access tokens that can be used to access protected resources.
The Spring Cloud Gateway is used as an OpenID client, that will authenticate users using Keycloak OIDC capabilities. Once the user is authenticated, Spring Cloud Gateway will use the Access Token received from Keycloak to forward requests to the appropriate microservice.
The microservices, as OpenID resource servers, will then validate the Access Token received from Spring Cloud Gateway to ensure that the user is authorized to access the requested resource.

\section{Discussion}

We developed the PHYLOViZ Web Platform, a user-friendly tool that  combines the functionalities of previous versions of PHYLOViZ, as well as integrates features from other tools like phyloDB and FLOWViZ. This integration creates a comprehensive and seamless user experience while ensuring the platform remains extensible and easy to maintain.

One of the key advantages of the proposed platform is its modular system architecture. This design allows for flexibility in adding new functionalities and components, ensuring that users have access to the latest advancements in phylogenetic analysis. Additionally, the modular architecture facilitates easy updates and enhancements to existing features, guaranteeing a cutting-edge platform for researchers.

To address the needs of large-scale analyses, the platform leverages server-side computation and parallel workflows. These computational capabilities enable efficient processing of large datasets, making it possible to handle large-scale phylogenetic analyses with ease. By optimizing computation resources, the platform ensures faster and more accurate results, ultimately enhancing the productivity of researchers.

Furthermore, the platform places a strong emphasis on reproducibility in phylogenetic research. To ensure reproducibility, the platform incorporates a comprehensive data management system that stores all data and computation results generated during the analysis process, as well as relevant metadata. This includes not only the raw input data but also workflow specifications, tool configurations and parameterizations, visualizations, transformations, and any intermediate steps undertaken.

As future work, we plan to expand the visual transformation capabilities, specifically by incorporating features like the collapse and expansion of parts of tree visualizations. Large phylogenetic trees can become complex and challenging to navigate, especially when dealing with a large number of branches or deep evolutionary relationships. By allowing users to selectively collapse or expand branches of the tree, the platform can provide a more manageable and focused view of the data. Moreover, we also plan to integration of additional analysis methods as well as tree layouts.

\section*{Acknowledgements}
The work reported in this article was supported by national funds through Fundação para a Ciência e a Tecnologia (FCT) with references DSAIPA/DS/0118/2020, UIDB/50021/2020 and LA/P/0078/2020.
\bibliographystyle{splncs04}
\bibliography{references}
\end{document}